\documentclass[aps, pra, floatfix, twocolumn, superscriptaddress, showpacs]{revtex4}

\usepackage{graphicx}
\usepackage{color}
\usepackage{calc}
\usepackage{array}
\usepackage{graphicx}
\usepackage{amsmath, amssymb}
\usepackage{natbib}
\usepackage{hyperref}
\usepackage{subfigure}

\def\d{\mathrm{d}}
\newcommand{\abs}[1]{\left \vert #1 \right \vert}
\newcommand{\ket}[1]{\left| #1 \right>} 
\DeclareMathOperator{\dd}{\mathrm{d}\!}
\DeclareMathOperator{\ii}{\mathrm{i}}

\begin{document}

\title{Macroscopic quantum superpositon states of\\
 two-component Bose-Einstein condensates}

\author{G{\'{a}}bor \surname{Csire}}%
\email{csire.gabor@wigner.bme.hu}
\affiliation{Institute of Physics,
             Budapest University of Technology and Economics, \\
             H-1111, Budafoki {\'{u}}t 8, Hungary}
\affiliation{Institute of Physics,
             E{\"{o}}tv{\"{o}}s Lor{\'{a}}nd University, \\
             H-1117, P{\'{a}}zm{\'{a}}ny P{\'{e}}ter s{\'{e}}t{\'{a}}ny 1/A, Hungary}

\author{Barnab{\'a}s Apagyi}
\email{apagyi@phy.bme.hu} \affiliation{Institute of Physics,
             Budapest University of Technology and Economics, \\
             H-1111, Budafoki {\'{u}}t 8, Hungary}

\date{\today}

\begin{abstract}

We examine a two-component Bose-Einstein condensate in a double-well potential.
We propose a model for the creation of many-particle macroscopic quantum superposition states.
The effect of dissipation on the formation of these states is also investigated
with the Monte-Carlo wavefunction technique.

\end{abstract}

\pacs{03.75.Gg, 03.75.Mn}

\maketitle

\section{Introduction \label{sec:introduction}}

The principle of superposition is at the heart of quantum mechanics.
Microscopic quantum superpositions are routinely observed in experiments,
but for macroscopic objects the decoherence time is almost instantaneous
~\cite{Zurek1991, Zurek2003}.
The famous thought experiment of Schr{\"{o}}dinger~\cite{Schrodinger1935} illustrates
the problems caused by the transition from quantum to classical regime in the
description of physical systems of increasing size.
At the quantum-classical boundary, systems should be isolated long enough,
so that creation of superposition of distinguishable states might be possible at
least on mesoscopic scale.
If quantum mechanics and classical mechanics are limiting cases
of the same dynamics, the theory, which describes the relation between them,
must be nonlinear and stochastic~\cite{Bassi2003}.
Nonlinearity is
crucial for the creation of superpositions of macroscopically distinguishable states,
the so called macroscopic quantum superposition states (MQSS).
These states were first suggested
by Yurke and Stoler~\cite{Yurke1986}. A small number of experimental realizations has been
reported.
Friedman et al.~\cite{Friedman2000} have shown that a SQUID can be put into a superposition
of two magnetic-flux states: one corresponding to a few microamperes of current flowing
clockwise, the other corresponding to the same amount of current flowing
anticlockwise.
Brune et al.~\cite{Brune1996, Raimond2001} produced superpositions
involving microwave photons.
MQSS with trapped ions have been created in Ref.~\cite{Leibfried2005}.
Ourjoumtsev et al.~\cite{Ourjoumtsev2007} also demonstrated experimental MQSS
using homodyne detection and photon number states as resources.

During the last decade ultracold gases became a very important testbed
for many predictions of condensed matter physics~\cite{Bloch2008}. The
underlying reason for this popularity is the high flexibility of the system
parameters: the strength of interaction, the type and strength of external
potential(s) the particles are moving in, or the phase difference between two
(or more) condensates. This versatility allows us to check and examine, for
example, predicted delicate phases in different dimensions, such as the
Bardeen-Cooper-Shrieffer \cite{Regal2004, Zwierlein2004} or the Mott-insulator
phase~\cite{Greiner2002} in 3D, the Berezinksii-Kosterlitz-Thouless
\cite{Hadzibabic2006, Schumayer2007} phase in 2D and the Tonks-Girardeau phase
in 1D~\cite{Paredes2004, Astrakharchik2005}.
With the advent of  Bose-Einstein condensates (BECs) a new experimental system has become available for
the investigation of macroscopic quantum superposition states.
Recently, several proposals for generation of MQSS in
 BECs
 have been reported
in the literature. Cirac et al.~\cite{Cirac1998} have provided a method
which involves adiabatic transfer of a two-species condensate to the many-body
ground state. Gordon and Savage~\cite{Gordon1999, Louis2001} use a two-component
BEC with two-body interactions and weak Josephson coupling between the
two components. In Ref.~\cite{Ruostekoski1998} a scheme of creating MQSS
with scattering light from two BECs moving with opposite velocities has been proposed.
MQSS of $\pi$-phase imprinted single component BEC have been predicted to exist
in a double-well potential~\cite{Mahmud2005}.

In this study we work out a quasi-1D Hamiltonian model which describes
two-component BEC states. In the two-mode approximation we solve
the eigenvalue problems and construct MQSS (cat states)
for small BEC systems appropriate for numerical treatment.
By investigating phase decoherence and time development we establish
signatures of cat states for larger BEC systems which develop into mixture states.

Section \ref{sec:model} deals with the model, in section \ref{sec:mqss} we
investigate MQSS in a double-well potential. Section \ref{sec:decoherence}
is left for treatment of decoherence and section \ref{sec:conclusion}
contains the conclusion and a summary.

\section{The model \label{sec:model}}

A two-component Bose-Einstein condensate can be described with
the following second quantized Hamiltonian at zero-temperature:

\begin{equation} \label{eq:ham}
H=H_1 + H_2 + H_{12},
\end{equation}
where
\begin{equation} \label{eq:ham_i}
\begin{split}
 H_i &=
 \int \d^3r \left(
 -\frac{\hbar^2}{2m_i} \Psi_i^\dag(\underline{r}) \Delta \Psi_i(\underline{r})+
     \Psi_i^\dag(\underline{r}) V_i(\underline{r}) \Psi_i(\underline{r}) \right) \\
 & + \frac{1}{2} g_{ii} \int \d^3r \,
 \Psi_i^\dag(\underline{r}) \Psi_i^\dag(\underline{r}) \Psi_i(\underline{r}) \Psi_i(\underline{r}),
\end{split}
\end{equation}
\begin{equation} \label{eq:ham_12}
 H_{12} = g_{12} \int \d^3r \,
 \Psi_1^\dag(\underline{r}) \Psi_2^\dag(\underline{r}) \Psi_1(\underline{r}) \Psi_2(\underline{r}).
\end{equation}
$V_i(\underline{r})$ is the external trapping potential. The coupling constants
$g_{ii}$ and $g_{12}$ are the intra- and inter-species atom-atom interaction strengths:
\begin{subequations}
\begin{eqnarray} \label{eq:g_ii}
 g_{ii} &=& \frac{4\pi\hbar^2}{m_i}a_{ii}, \\
 \label{eq:g_12}
 g_{12} &=& \frac{2\pi\hbar^2}{m_{12}}a_{12},
\end{eqnarray}
\end{subequations}
where $m_{12}$ denotes the reduced mass
\begin{equation} \label{eq:red_mass}
m_{12} = \frac{m_1m_2}{m_1+m_2},
\end{equation}
and $a_{ij}$ is the s-wave scattering length which can be tuned via Feshbach resonance.
$\Psi_i^\dag(\underline{r})$ and $\Psi_i(\underline{r})$ are the bosonic
creation and annihilation operators for the two species satisfying the usual
commutation rules and normalized to the number $N_i$ of particles of species $i$.

We will assume that the external potentials have the following form
\begin{equation} \label{eq:3D-HO}
V_i(\underline{r})={1\over 2}m_i\omega_{i,\bot}^2(y^2+z^2) + V_i(x),
\end{equation}
where
\begin{equation} \label{eq:DW-x}
 V_i(x)={1\over 2}m_i\omega_{i,x}^2\left(x^2+A\mathrm{e}^{-Bx^2}\right).
\end{equation}
We define
\begin{eqnarray}
\label{eq:a_L-R}
a^L_i=\frac{a^{(0)}_i + a^{(1)}_i}{\sqrt{2}}, \qquad
a^R_i=\frac{a^{(0)}_i - a^{(1)}_i}{\sqrt{2}}, \\
\label{eq:phi_L-R}
{\phi}^L_i=\frac{{\phi}^{(0)}_i + {\phi}^{(1)}_i}{\sqrt{2}}, \qquad
{\phi}^R_i=\frac{{\phi}^{(0)}_i - {\phi}^{(1)}_i}{\sqrt{2}},
\end{eqnarray}
where ${\phi}^{(0)}_i$ and ${\phi}^{(1)}_i$ are the
mean-field ground and first excited states for an atom of species $i$,
and they are normalized to one.
$a^{(0)}_i$ and  $a^{(1)}$ are the annihilation operators
in the ground and excited state, respectively and satisfy
the usual commutation relations.
${\phi}^{(0)}_i$ and ${\phi}^{(1)}_i$
can be obtained from the coupled
Gross-Pitaevskii (GP) equations \cite{Gross1961, Pitaevskii1961}.
The confinement of the potential is relatively strong in the radial
directions, therefore, one may assume that the evolution in the weak
x-direction decouples from that of in the strong [yz]-plane.
Using these simplifying assumptions the three-dimensional
equation can be reduced to the following coupled, time-independent,
one-dimensional equations~\cite{Csire2010}
\begin{subequations}
\begin{equation} \label{eq:coupled_GP1}
   \mu_1 {\phi}_{1}
   =
   \left \lbrack
      - \frac{1}{2} \, \partial_{xx}
      + \widetilde{V}_1(x)
      + {b_{11} \abs{{\phi}_{1}}^{2}}
      + {b_{12} \abs{{\phi}_{2}}^{2}}
   \right \rbrack \! {\phi}_{1},
\end{equation}
\begin{equation} \label{eq:coupled_GP2}
   \mu_2 {\phi}_{2}
   =
   \left \lbrack
      - \frac{\kappa}{2} \, \partial_{xx}
      +   \frac{1}{\kappa}\widetilde{V}_2(x)
      + {b_{21} \abs{{\phi}_{1}}^{2}}
      + {b_{22} \abs{{\phi}_{2}}^{2}}
   \right \rbrack \! {\phi}_{2},
\end{equation}
\end{subequations}
where $\mu_1$ and $\mu_2$ are the chemical potentials and
\begin{equation} \label{eq:widetilde_V}
 \widetilde{V}_i(x)=\frac{\lambda_i^2}{2} \left( x^2 +
         \widetilde{A}\mathrm{e}^{-\widetilde{B} x^2} \right),
\end{equation}
$b_{11}=2a_{11}N_1,$ $b_{22}=2a_{22} N_1\kappa/\gamma,$
$b_{12}=b_{21}=2a_{12}N_1(1+\kappa)/(1+\gamma),$
$\gamma=\omega_{2,\bot}/\omega_{1,\bot},$ $\kappa=m_1/m_2,$
$\lambda_{1} = \omega_{1,x} / \omega_{1,\bot}$,
$\lambda_{2} = \omega_{2,x} / \omega_{1,\bot}$,
$\widetilde{A}=A/a_0^2$, and $\widetilde{B}=B \, a_0^2$.
The distances are measured
in units $a_0=\sqrt{{\hbar}/{m_1\omega_{1,\bot}}}$.
The normalization is
such that $\int|\phi_1|^2 \dd x=1$ and $\int|\phi_2|^2 \dd x=N_2/N_1$.
Moreover, the relation $\gamma^2=\kappa$ must hold if both species
experience the same HO potential.

According to the two-mode approximation~\cite{Smerzi1997}
we expand the field operator as follows
\begin{equation} \label{eq:two_mode}
 \Psi_i=\left[a^{L}_i {\phi}^{L}_i(x) + a^{R}_i {\phi}^{R}_i(x)\right]\chi_{i,\bot}(y,z),
\end{equation}
where
$\chi_{i,\bot}$ satisfies the corresponding 2D HO Schr\"odinger
equation. It is convenient to introduce angular momentum operators~\cite{Micheli2003}
\begin{subequations}
\begin{eqnarray}
\label{eq:J_x}
J_i^x &=& \frac{1}{2} \left[ \left(a_i^R\right)^\dag a_i^L + \left(a_i^L\right)^\dag a_i^R \right], \\
\label{eq:J_y}
J_i^y &=& \frac{-\mathrm{i}}{2} \left[
          \left(a_i^R\right)^\dag a_i^L - \left(a_i^L\right)^\dag a_i^R \right], \\
\label{eq:J_z}
J_i^z &=& \frac{1}{2} \left[ \left(a_i^R\right)^\dag a_i^R - \left(a_i^L\right)^\dag a_i^L \right],
\end{eqnarray}
\end{subequations}
where $J_i^x$, $J_i^y$ correspond to the correlation between the two wells and $J_i^z$ is the particle number difference between the two wells for the species $i$.
Furthermore, $J_i^2$ is the Casimir invariant of the SU(2) algebra:
\begin{equation} \label{eq:J^2}
 J_i^2 = \left(J_i^x\right)^2 + \left(J_i^y\right)^2 + \left(J_i^z\right)^2= \frac{N_i}{2}\left(\frac{N_i}{2}+1 \right).
\end{equation}

By substituting (\ref{eq:two_mode}) into (\ref{eq:ham}) gives the following
result within the improved two-mode approximation~\cite{Ananikian2006, Satija2009}
\begin{equation} \label{eq:hamilton}
\begin{split}
 H &= \sum_{i=1}^2 \left [
    \left(\alpha_i+ \epsilon_i\right) J_i^x + \beta_i \left(J_i^x\right)^2
    + \gamma_i \left(J_i^z\right)^2 \right]\\
    &+ \alpha_{12} J_1^x  + \alpha_{21} J_2^x
    + \beta_{12} J_1^x J_2^x
    + \gamma_{12} J_1^z J_2^z,
\end{split}
\end{equation}
where
\begin{widetext}
\begin{subequations}
\begin{eqnarray}
\label{eq:epsilon}
 \epsilon_i &=& -\frac{2}{a_0}\int \dd {x} \left[
           \frac{\hbar^2}{2m}\left(\nabla \phi_i^{L} \right)\left(\nabla \phi_i^{R} \right) +
            (V(x) + \hbar\omega_{i,\bot})\phi_i^{L}\phi_i^{R} \right],\\
\label{eq:alpha_i}
 \alpha_i &=& (N_i-1)g_{ii} \frac{m_i\omega_{i,\bot}}{2a_0 \pi\hbar}
              \int \dd {x} \left[ \left(\phi_i^{R}\right)^3 \phi_i^{L} +
              \left(\phi_i^{L}\right)^3 \phi_i^{R}\right],\\
\label{eq:alpha_ij}
\alpha_{ij} &=& -\frac{2}{a_0\pi\hbar}\frac{N_j}{N_i}g_{12}
              \frac{m_1\omega_{1,\bot}m_2\omega_{2,\bot}}{m_1\omega_{1,\bot}+m_2\omega_{2,\bot}}
              \int \dd {x}
              \left[ \left(\phi_i^{R}\right)^2 \phi_j^{R}\phi_j^{L} +
              \left(\phi_i^{L}\right)^2 \phi_j^{R}\phi_j^{L}\right],\\
\label{eq:beta_i}
\beta_i &=& g_{ii} \frac{m_i\omega_{i,\bot}}{a_0\pi\hbar}
            \int \dd {x} \left(\phi_i^{L}\right)^2 \left(\phi_i^{R}\right)^2,\\
\label{eq:beta_12}
\beta_{12} &=&  \frac{4g_{12}}{a_0\pi\hbar}
              \frac{m_1\omega_{1,\bot}m_2\omega_{2,\bot}}{m_1\omega_{1,\bot}+m_2\omega_{2,\bot}}
               \int \dd {x} \, \phi_1^{L}\phi_1^{R}\phi_2^{L}\phi_2^{R},\\
\label{eq:gamma_i}
\gamma_i &=& g_{ii} \frac{m_i\omega_{i,\bot}}{8a_0\pi\hbar}
             \int \dd {x} \left[\left(\phi_i^{L}\right)^2-\left(\phi_i^{R}\right)^2\right]^2,\\
\label{eq:gamma_12}
\gamma_{12} &=& \frac{g_{12}}{a_0\pi\hbar}
               \frac{m_1\omega_{1,\bot}m_2\omega_{2,\bot}}{m_1\omega_{1,\bot}+m_2\omega_{2,\bot}}
               \int \dd {x} \left[\left(\phi_1^{L}\right)^2-\left(\phi_1^{R}\right)^2\right]
                 \left[\left(\phi_2^{L}\right)^2-\left(\phi_2^{R}\right)^2\right].
\end{eqnarray}
\end{subequations}
\end{widetext}
We measure the mass in units of the unified atomic mass unit.

A similar model was introduced within a simple two-mode approach
in Ref.~\cite{Mazzarella2010}.

\section{Macroscopic quantum superpositon states \label{sec:mqss}}

In this section we want to determine the domain of $a_{ij}$'s where it is possible
 to create cat states.
We solve the eigenvalue problem of the Hamiltonian \eqref{eq:hamilton}.
The state space can be spanned in the $(N_1+1)(N_2+1)$ Fock state basis
$\ket{n_1^L,N_1-n_1^L}_1 \ket{n_2^L,N_2-n_2^L}_2$, where
$n_i^L$  is the number of particles of species $i$ in the left well.
An arbitrary state reads as follows
\begin{equation} \label{eq:Phi}
\ket{\Phi} = \sum_{n_1^L=0}^{N_1} \sum_{n_2^L=0}^{N_2} c_{n_1^L,n_2^L}
             \ket{n_1^L,N_1-n_1^L}_1 \ket{n_2^L,N_2-n_2^L}_2,
\end{equation}
where
\begin{equation} \label{eq:basis}
\begin{split}
 &\ket{n_1^L,N_1-n_1^L}_1 \ket{n_2^L,N_2-n_2^L}_2 =\\
 &\left[ \prod_{i=1}^2
 \frac{\left(\left(a_i^L\right)^\dag\right)^{n_i^L}}{\sqrt{n_i^L!}}
 \frac{\left(\left(a_i^R\right)^\dag\right)^{N_i-n_i^L}}{\sqrt{\left(N_i-n_i^L\right)!}}
 \right]
 \ket{0,0}_1 \ket{0,0}_2.
\end{split}
\end{equation}
To obtain the parameters appearing in the Hamiltonian \eqref{eq:hamilton}  we need to determine the mean-field ground and first excited states
 ${\phi}^{(0)}_i$ and ${\phi}^{(1)}_i$
 from the time-independent coupled Gross-Pitaevskii equations
\eqref{eq:coupled_GP1}-\eqref{eq:coupled_GP2}. This can be done in case of the ground state ${\phi}^{(0)}_i$   by using the imaginary time
evolution method~\cite{Lehtovaara2007} combined with the split-step
operator technique~\cite{Javanainen2006}. For getting the first excited state ${\phi}^{(1)}_i$ we use the shooting method~\cite{numrec}.

We define $P$ as the number $N_{cat}$ of superpositions of two distinguishable states (associated with the cat states or MQSS) divided by
the total number $N_{tot}=(N_1+1)(N_2+1)$ of eigenstates ordered in sequence of growing eigenenergies,
\begin{equation} \label{eq:P}
P=\frac{N_{cat}}{N_{tot}}.
\end{equation}
In practice the selection of cat states proceeds via the search of the $N_{tot}$ states
(obtained by diagonalisation of \eqref{eq:hamilton}) for those members which possess
two expansion coefficients $|c_{n_1^L,n_2^L}|^2$ greater than 0.4.

For our numerical investigation we fix the following parameters:
$m_1=m_2=87$,
$\omega_{1,\bot}=\omega_{2,\bot}=2\pi\times 710$~Hz,
$\lambda_1=\lambda_2=0.2$,
$\widetilde{A}=50$,
$\widetilde{B}=10$.
This means that our exploratory investigation for two-component cat states will be carried out for the case of a quasi one-dimensional (cigar-like) $^{87}$Rb Bose gas where the two components correspond to two hyperfine states of the rubidium atom. Such two-component BEC systems are routinely  produced in laboratories~\cite{Papp2008} with particle numbers $N_1\sim N_2\sim 10000$. But simulating such a great number of atoms in the gas is beyond the capacity of today's  computers.
We think, however, that features of cat states to be presented here for lower
particle numbers $N_1$ and $N_2$ will be valid also for larger $N_i$ values.
Therefore we shall carry out the calculations at particle numbers $N_1$, $N_2\sim 1-20$ and vary the scattering lengths in the possible domain which can be achieved by using the Feshbach resonance method~\cite{Chin2010}.

\begin{figure}[hbt!]
   \includegraphics[width=\textwidth/7*3]{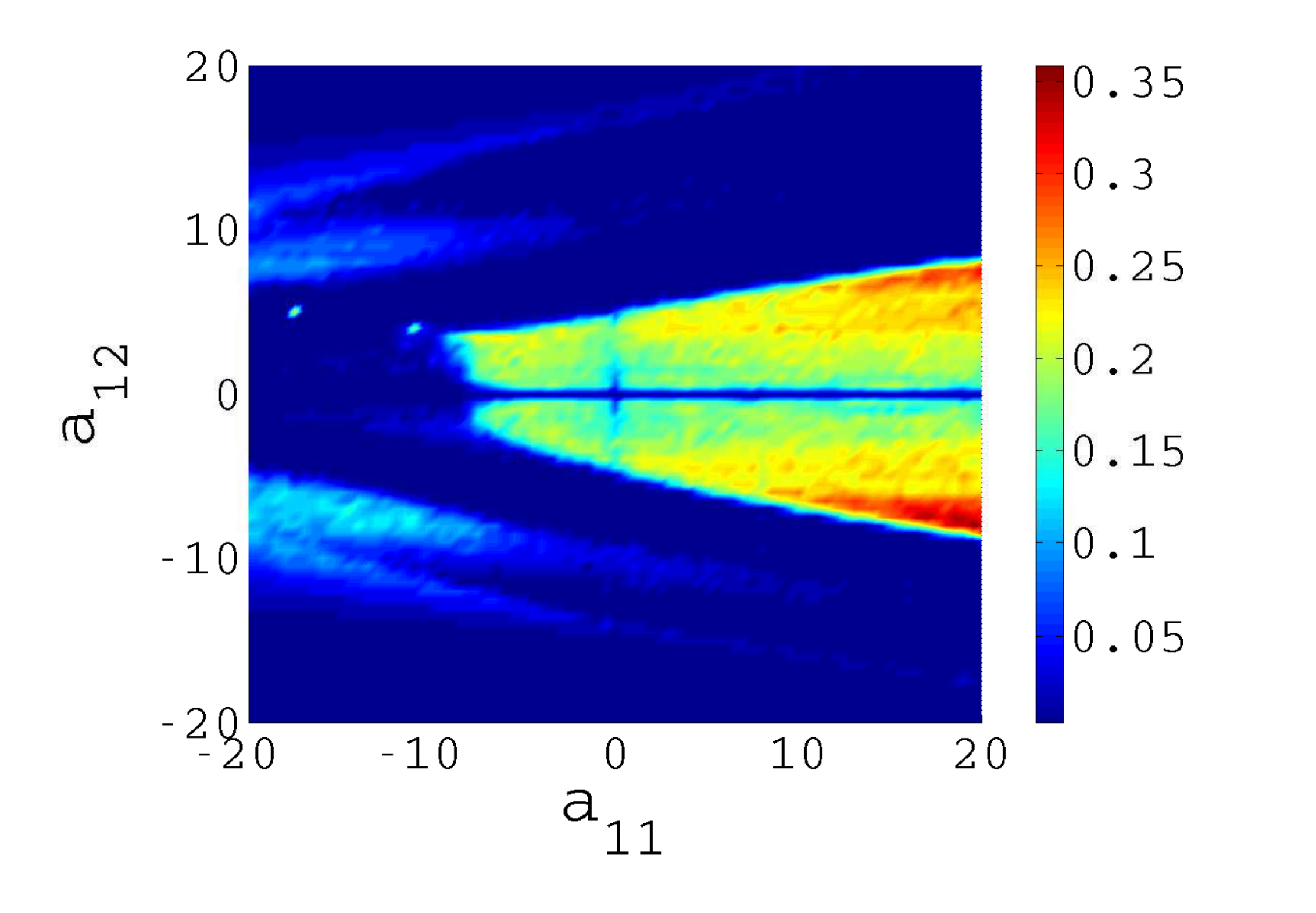}
   \caption{\label{fig:prob1}(Color online) $P$ as function of $a_{11}$ and $a_{12}$ in nm;
		fixed parameters: $N_1=4$, $N_2=30$, $a_{22}=2$~nm.}
\end{figure}

In Fig.~\ref{fig:prob1}  the probability $P$ of cat state formation is shown as function of $a_{11}$ and $a_{12}$ with fixed parameters $a_{22}=2$ nm, and particle  numbers $N_1=4$ and $N_2=30$. We see an almost symmetrical triangle which is cut at $a_{12}=0$ where there is no two-component cat states. This is reasonable because the components are decoupled and, at the most, one-component cat states can be formed with such parameter values. We observed however also that there is no two-component cat states when the sign of $a_{22}$ is changed to negative
(at the above fixed parameters).

Therefore, it is reasonable to study the formation of cat states as function of the particle numbers $N_1$, $N_2$.

\begin{figure}[hbt!]
   \includegraphics[width=\textwidth/7*3]{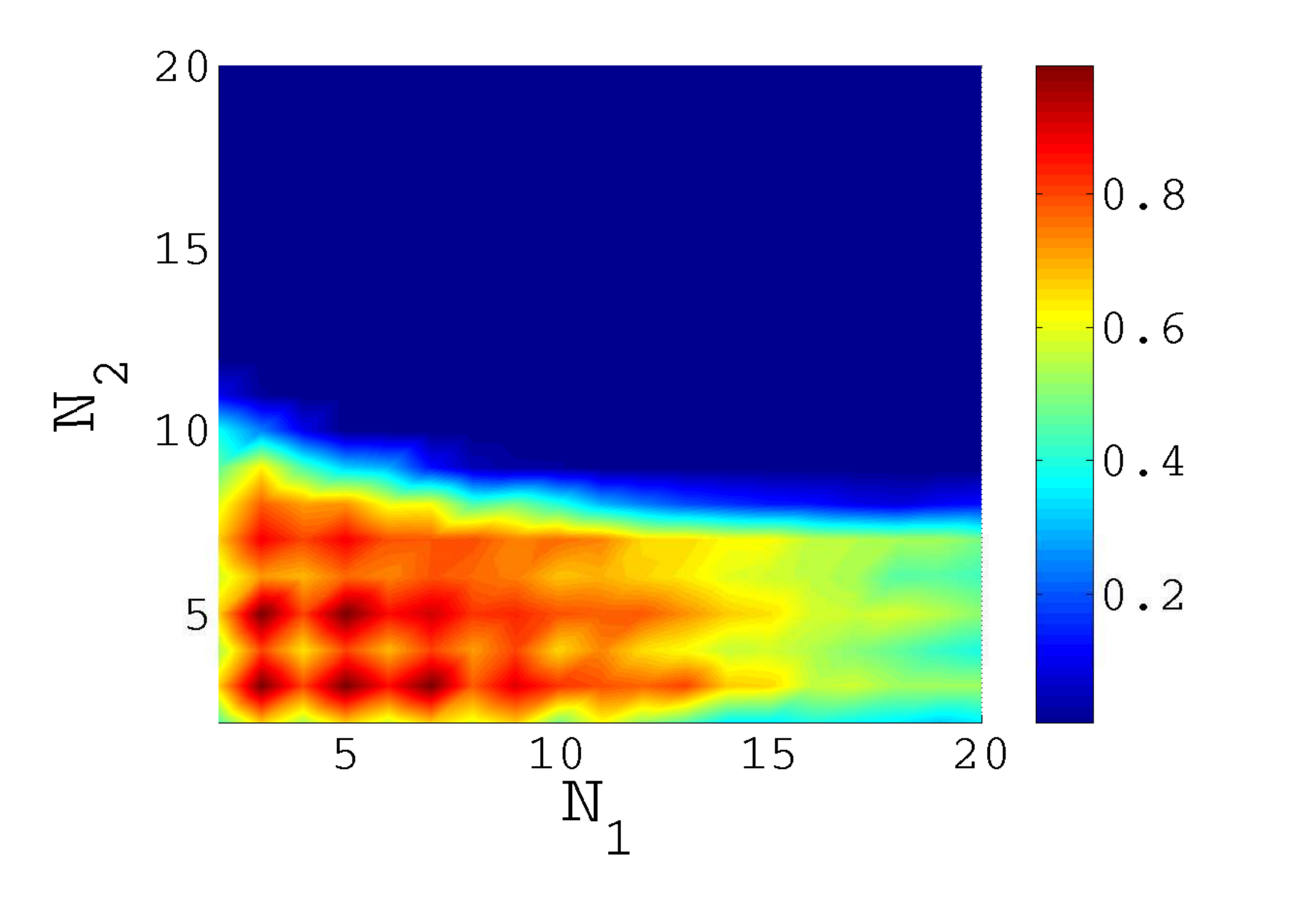}
   \caption{\label{fig:prob2}(Color online) $P$ as function of $N_1$ and $N_2$;
		fixed parameters: $a_{11}=2$~nm, $a_{12}=4$~nm, $a_{22}=-2$~nm.}
\end{figure}

In Fig.~\ref{fig:prob2}  $P$ is shown as function of $N_1$ and $N_2$ at positive $a_{11},a_{12}$ and negative $a_{22}$ values. A sharp horizontal border can be observed from $N_2 > 10$, independent of $N_1$. This result explains the former finding of no cat state when $a_{22}$ has been changed to negative values. We numerically observed also that this horizontal phase separation border (between cat states and no cat states) can be raised to larger $N_2$ values by raising also the value of $a_{22}$.

\begin{figure}[hbt!]
\subfigure[]{\label{fig:prob3}\includegraphics[width=\textwidth/7*3]{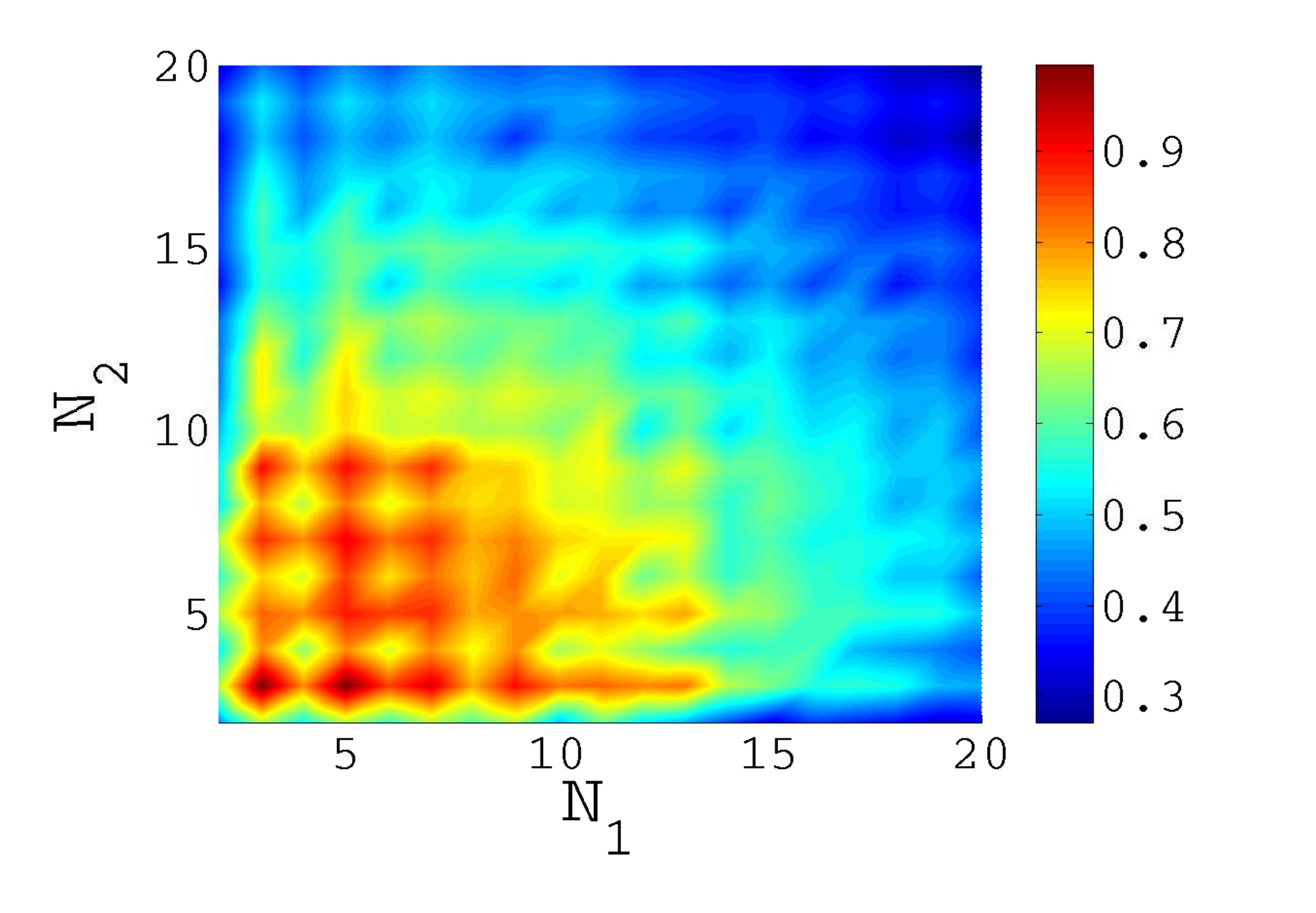}}
\\
\subfigure[]{\label{fig:prob3b}\includegraphics[width=\textwidth/7*3]{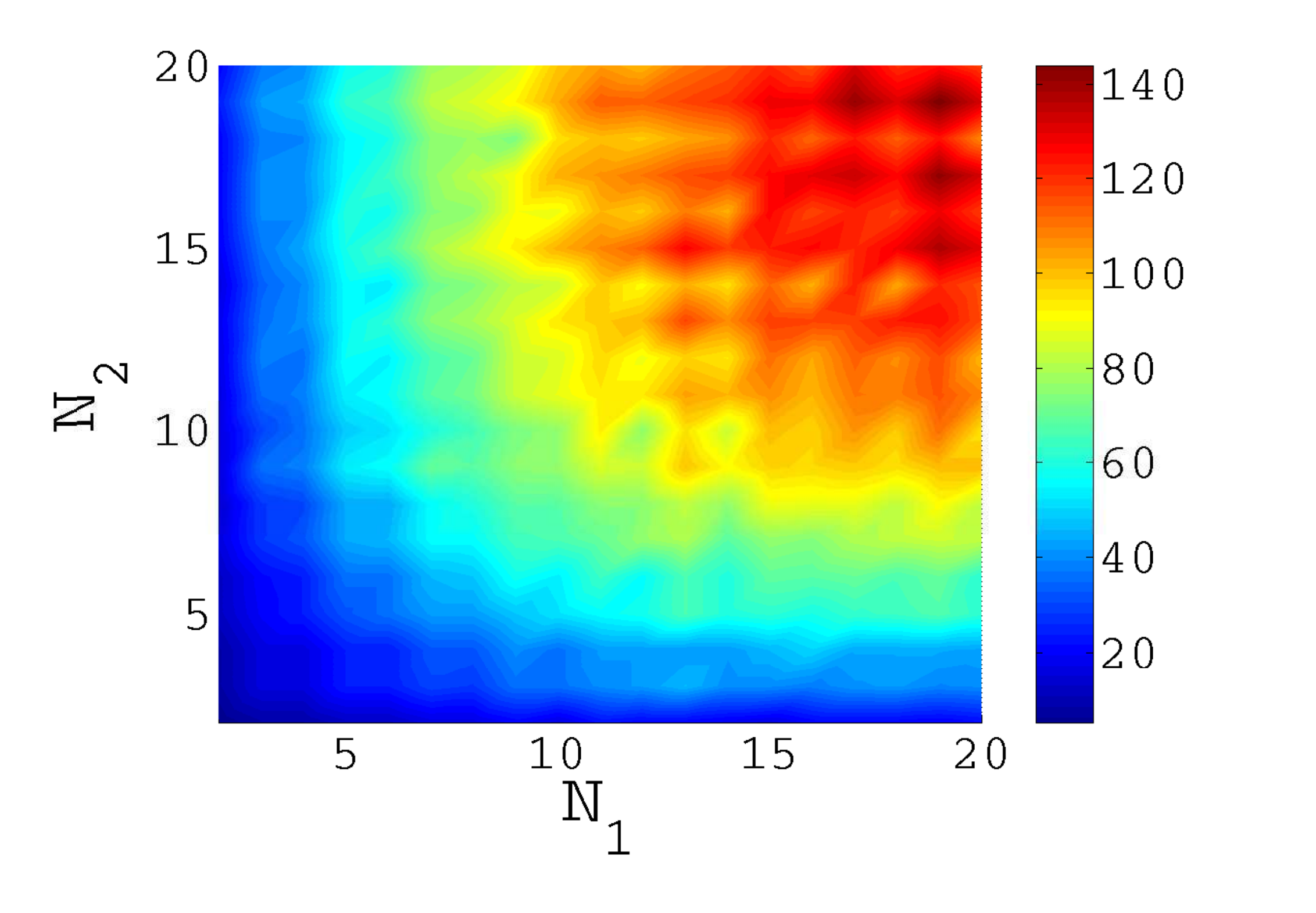}}
\caption{(Color online) Fixed parameters: $a_{11}=2$~nm, $a_{12}=4$~nm, $a_{22}=3$~nm.
(a) $P$ as function of $N_1$ and $N_2$.
(b) $N_{cat}$ as function of $N_1$ and $N_2$.}
\end{figure}

In Fig.~\ref{fig:prob3} all the scattering lengths have been fixed at positive values. It can be seen that the probability $P$ of forming cat states is lowered at growing particle numbers (decline is slower for component with smaller scattering length). But this means only that greater particle numbers involve more $N_{tot}$  states and thus a smaller fraction becomes cat states. This can be visualized if the result of the same calculation is shown not for the probability $P$ but for the number $N_{cat}$  of cat states. In Fig.~\ref{fig:prob3b} we see that the number of cat states are growing up to 140 at particle numbers $N_1= N_2=20$ which means however total number of states $N_{tot}=441$, and therefore gives for $P$ only the value 0.3 which has been represented by blue color in Fig.~\ref{fig:prob3}.

\begin{figure}[hbt!]
\subfigure []{\label{fig:prob4}\includegraphics[width=\textwidth/7*3]{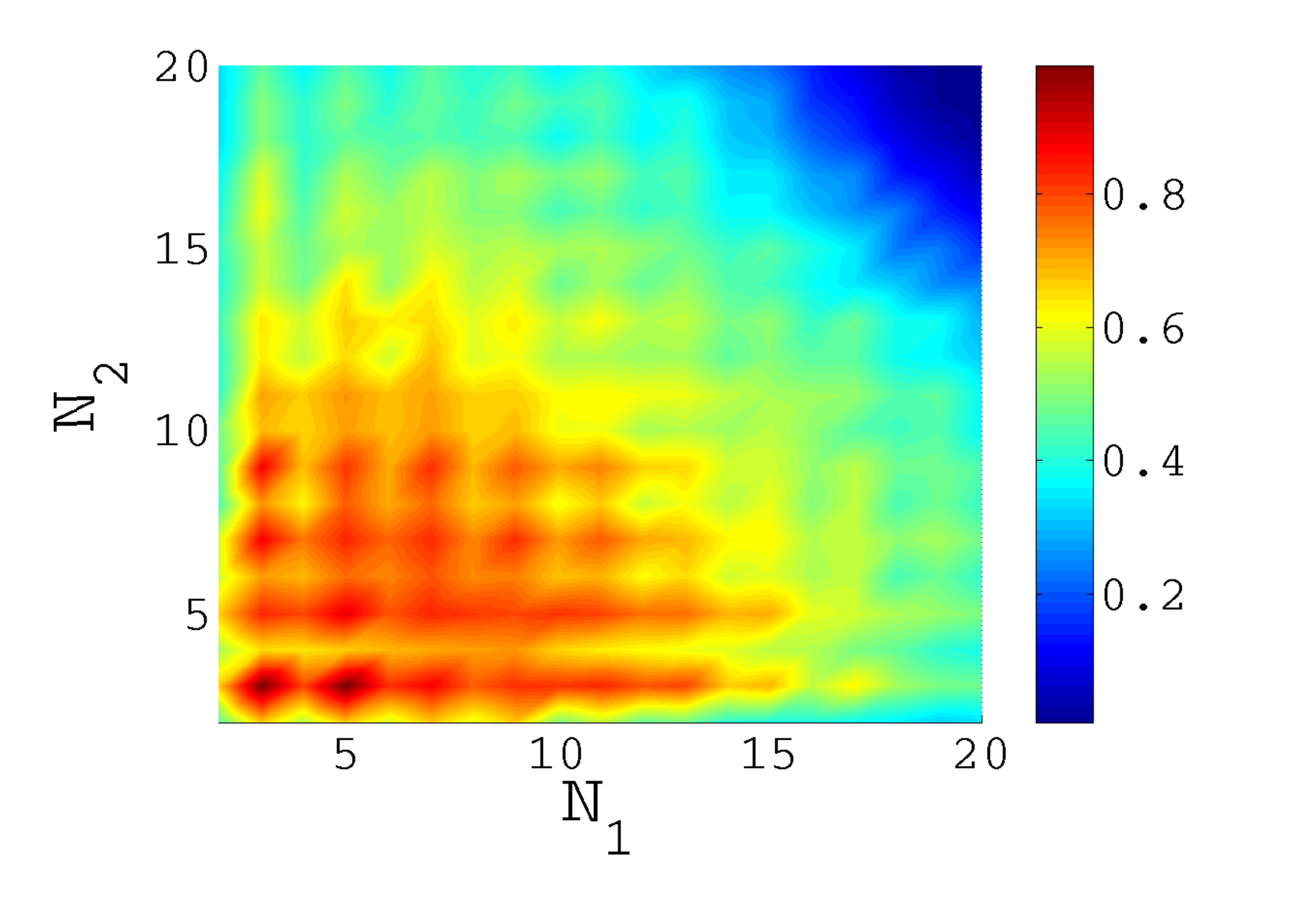}}
\\
\subfigure[]{\label{fig:prob4b}\includegraphics[width=\textwidth/7*3]{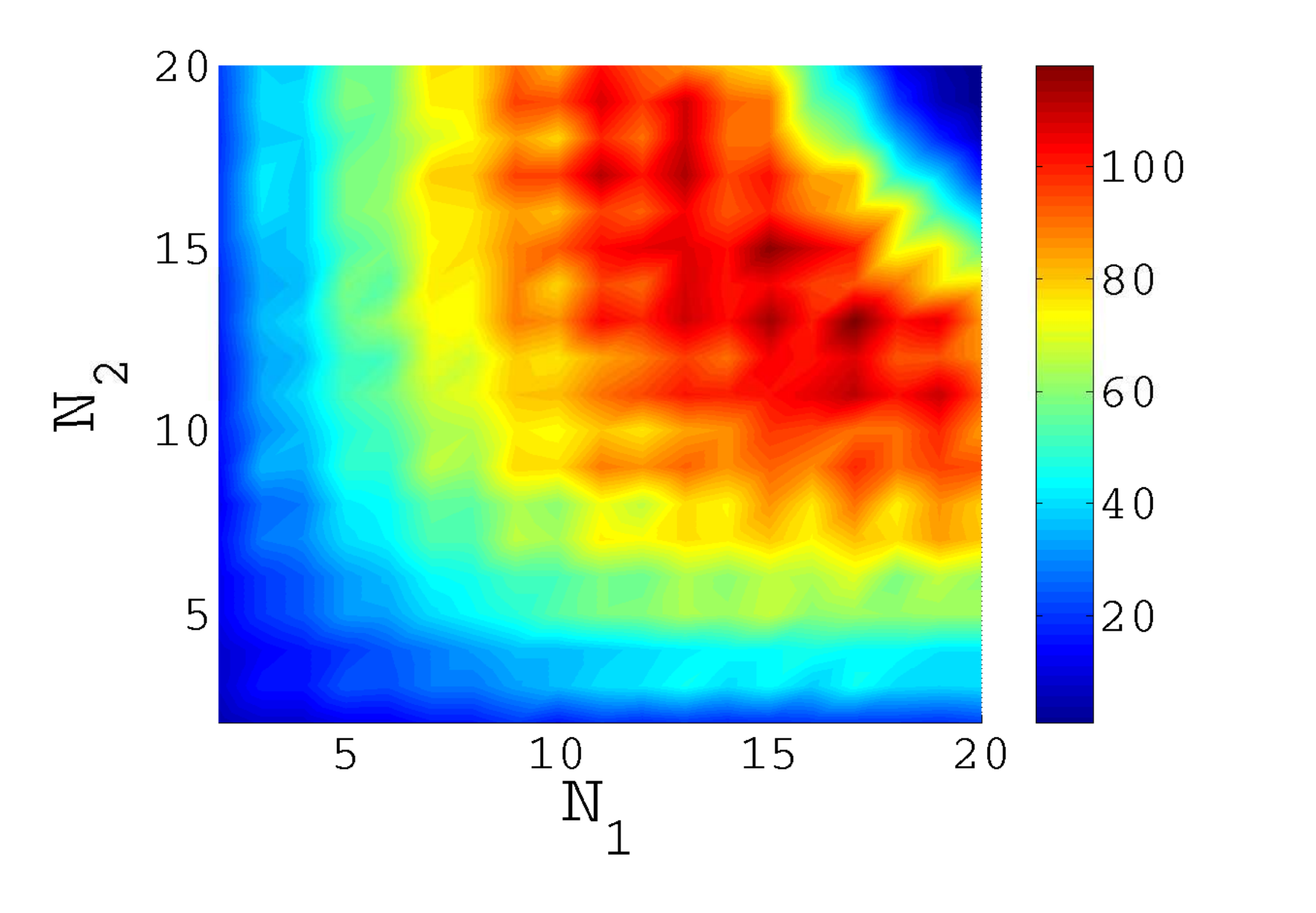}}
\caption{(Color online) Fixed parameters:  $a_{11}=2$~nm, $a_{12}=-4$~nm, $a_{22}=3$~nm.
(a) $P$ as function of $N_1$ and $N_2$.
(b) $N_{cat}$ as function of $N_1$ and $N_2$.}
\end{figure}

In Fig.~\ref{fig:prob4} the intra-species scattering lengths $a_{11},a_{22}$ are positive while the inter-species scattering length $a_{12}$ is negative.  According to the color bars we see that $P$ decreases here more strongly compared to Fig.~\ref{fig:prob3} but the number of cat states is also decreasing, as can be seen in Fig.~\ref{fig:prob4b} where $N_{cat}$ is visualized for the same set of parameters.

\begin{figure}[hbt!]
\subfigure []{\label{fig:prob5}\includegraphics[width=\textwidth/7*3]{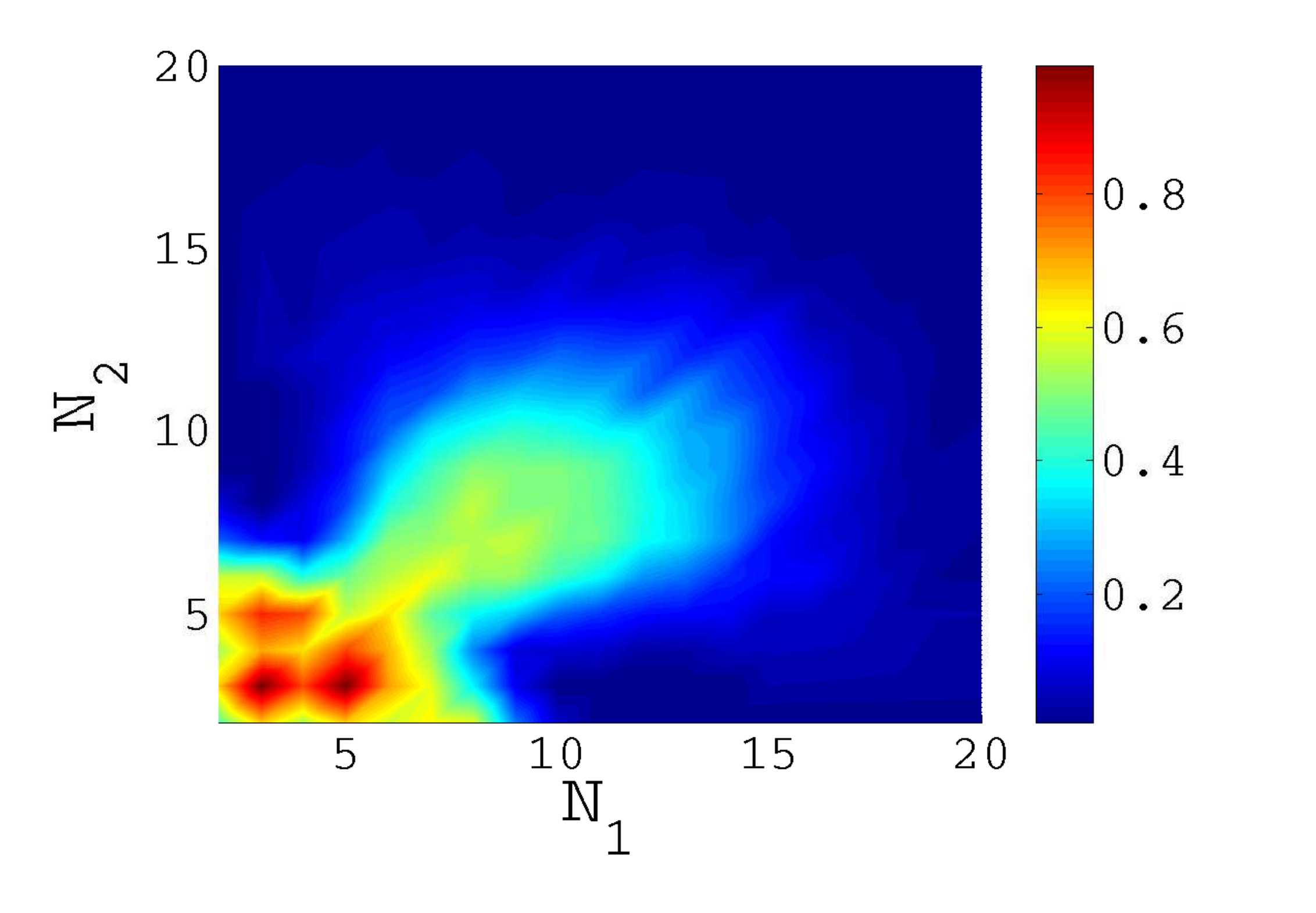}}
\\
\subfigure[]{\label{fig:prob5b}\includegraphics[width=\textwidth/7*3]{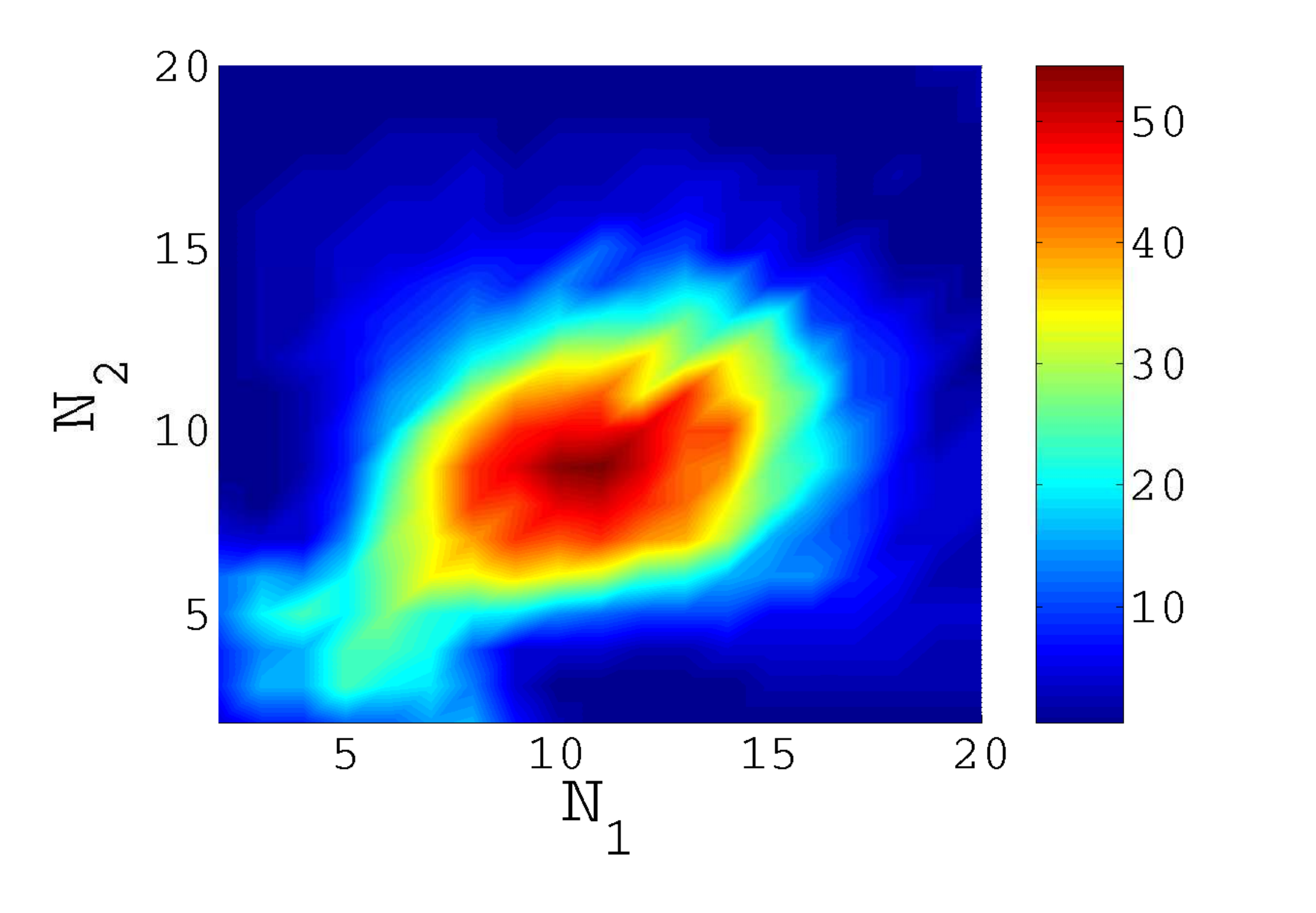}}
\caption{(Color online) Fixed parameters: $a_{11}=-2$~nm, $a_{12}=-4$~nm, $a_{22}=-3$~nm.
(a) $P$ as function of $N_1$ and $N_2$.
(b) $N_{cat}$ as function of $N_1$ and $N_2$.}
\end{figure}

In Fig.~\ref{fig:prob5} and Fig.~\ref{fig:prob5b} all scattering lengths take negative values. We see that for larger particle numbers there are no cat states at all.

\section{Decoherence \label{sec:decoherence}}

We study the effect of dissipation. The dominant source is the  thermal
cloud of noncondensed atoms. Other possible sources are ambient magnetic fields,
different scattering lengths, three-body losses~\cite{Dalvit2000}.
We will only investigate the effect of
elastic two-body interactions between the noncondensed and condensed
atoms. This type of interaction preserves the number of condensed atoms and
leads to phase-damping which dominates for MQSS~\cite{Louis2001}.

The most general quantum mechanical Markovian master equation has the
form~\cite{Lindblad1976}
\begin{equation} \label{eq:master_eq}
 \frac{\partial}{\partial t}{\rho}= -\frac{\ii}{\hbar}[{H},\widehat{\rho}] +
\sum_{\alpha} \left[ \mathcal{L}_{\alpha} {\rho}\mathcal{L}_{\alpha}^\dagger
- \frac{1}{2}\left\{ {\mathcal{L}}_{\alpha}^\dagger \mathcal{L}_{\alpha},{\rho} \right\} \right].
\end{equation}
where $\rho$ is the density matrix of system states, $\mathcal{L}_{\alpha}$'s are
the Lindblad operators containing system-environmental interaction, and
$\{\cdot,\cdot\}$ is the anticommutator.
The Lindblad formalism is valid when the coupling between
the system and environment is weak
~\cite{Talkner1986}. We use the following
Lindblad generators to describe the phase-damping:
\begin{subequations}
\begin{eqnarray}
\label{eq:L_1}
\mathcal{L}_{1}=\sqrt{\kappa_1} n_1^L, \\
\label{eq:L_2}
\mathcal{L}_{2}=\sqrt{\kappa_1} n_1^R, \\
\label{eq:L_3}
\mathcal{L}_{3}=\sqrt{\kappa_2} n_2^L, \\
\label{eq:L_4}
\mathcal{L}_{4}=\sqrt{\kappa_2} n_2^R,
\end{eqnarray}
\end{subequations}
where $\kappa_1$ and $\kappa_2$ are the phase-damping
rates, which are proportional to the number density of noncondensed atoms.

The dynamics of the master equation is studied in terms of stochastic
trajectories of state vectors~\cite{Molmer1993, Steinbach1995}. In this
approximation the density matrix of the system is written as the
average of the stochastic wavefunctions:
\begin{equation} \label{eq:rho_MC}
 \rho(t)=\frac{1}{MC}\sum_{i=1}^{MC}\left|\Phi^{(i)}(t)\right>\left<\Phi^{(i)}(t)\right|,
\end{equation}
where $\left|\Phi^{(i)}(t)\right>$ is the $i$th stochastic
trajectory of the state vector, and $MC$ is the number of the simulations.
In the evolution of the system two distinct elements can eventuate
for small $\Delta t$. Either one of the quantum jumps occurs (with the probablity
$\dd p_{\alpha}= \left< \mathcal{L}_{\alpha}^\dagger \mathcal{L}_{\alpha}\right>\Delta t$) which
can be represented as an action of the corresponding Lindblad operator on the state of the system
\begin{equation} \label{eq:qm_jump}
 \left|\Phi\right>\longmapsto  \mathcal{L}_{\alpha} \left|\Phi\right>,
\end{equation}
or the wavefunction is evolved by a non-Hermitian effective Hamiltonian (with the probablity
$1-\sum_{\alpha}\dd p_{\alpha}$), which has the form:
\begin{equation} \label{eq:H_eff}
 H_{eff}=H-\frac{\mathrm{i}\hbar}{2}\sum_{k=1}^{4} \mathcal{L}_k^{\dagger}\mathcal{L}_k.
\end{equation}
The trajectories must be normalized after every time step.
The Monte-Carlo approach has many advantages. Calculating a set of trajectories is more
economical, than solving the master equation with a Runge-Kutta method.
Moreover, a single-stochastic trajectory can be interpreted as a single realization of
an experiment manipulating a unique quantum system~\cite{Bergquist1986}.

Our calculation will be performed with parameter set $a_{11}=2$~nm, $a_{12}=4$~nm, $a_{22}=3$nm, $N_1=5$, $N_2=20$ and phase damping rates $\kappa_1=\kappa_2=0.005$. The density matrix is obtained
by averaging $MC=5000$ trajectories.
Initially, among the 126 eigenstates there are about 60 cat states as Fig.~\ref{fig:prob3b} suggests. From these we shall investigate only two which can be written to a good approximation as
$(\ket{1,4}_1\ket{20,0}_2+\ket{4,1}_1\ket{0,20}_2)/\sqrt{2}$ (called the 61th excited state) and
$(\ket{2,3}_1\ket{7,13}_2+\ket{3,2}_1\ket{13,7}_2)/\sqrt{2}$ (the 63th excited state) using the standard notation of the expansion \eqref{eq:Phi}.

\begin{figure}[hbt!]
   \subfigure[]{\includegraphics[width=\textwidth/7*3]{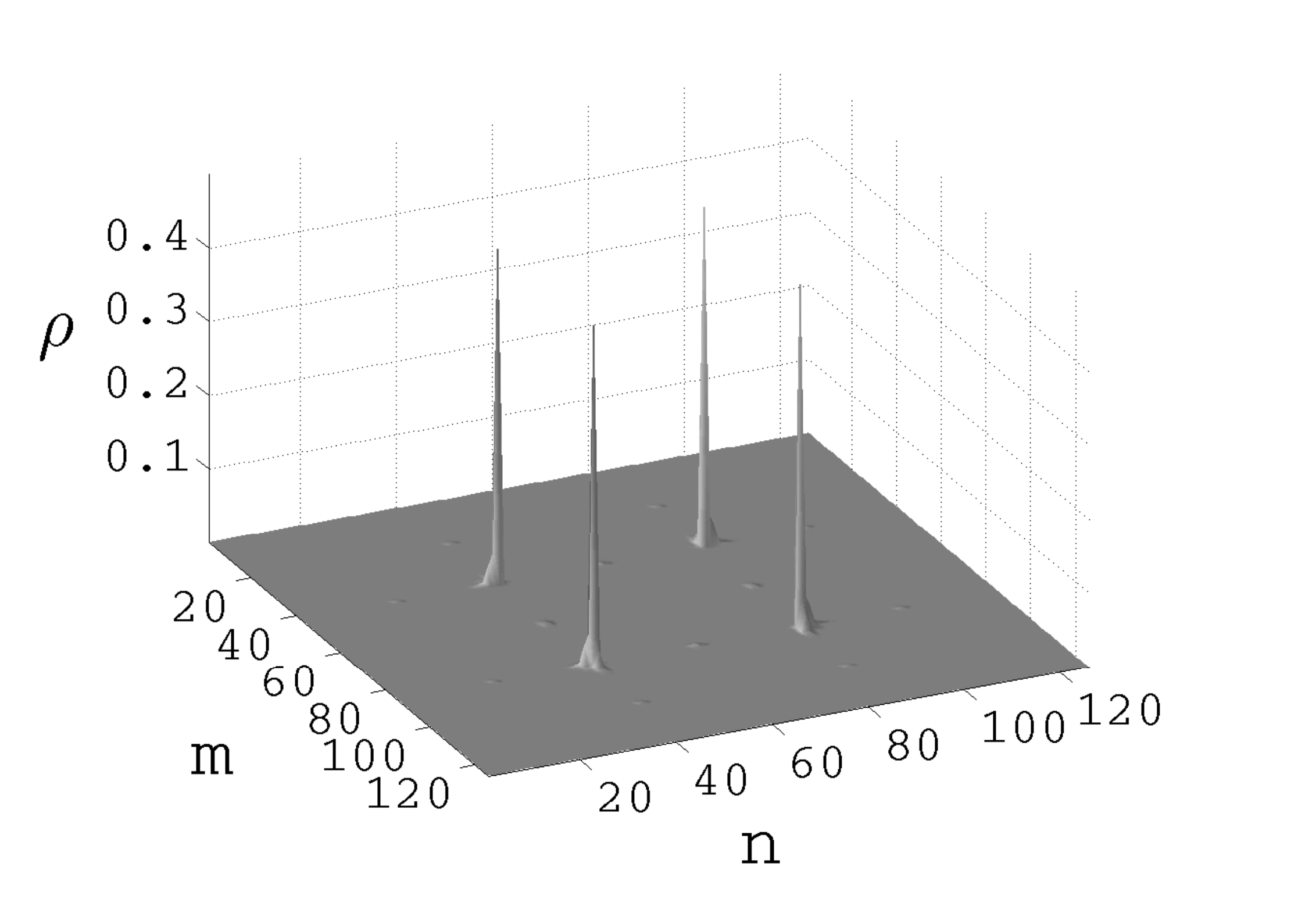}}\\
   \subfigure[]{\includegraphics[width=\textwidth/7*3]{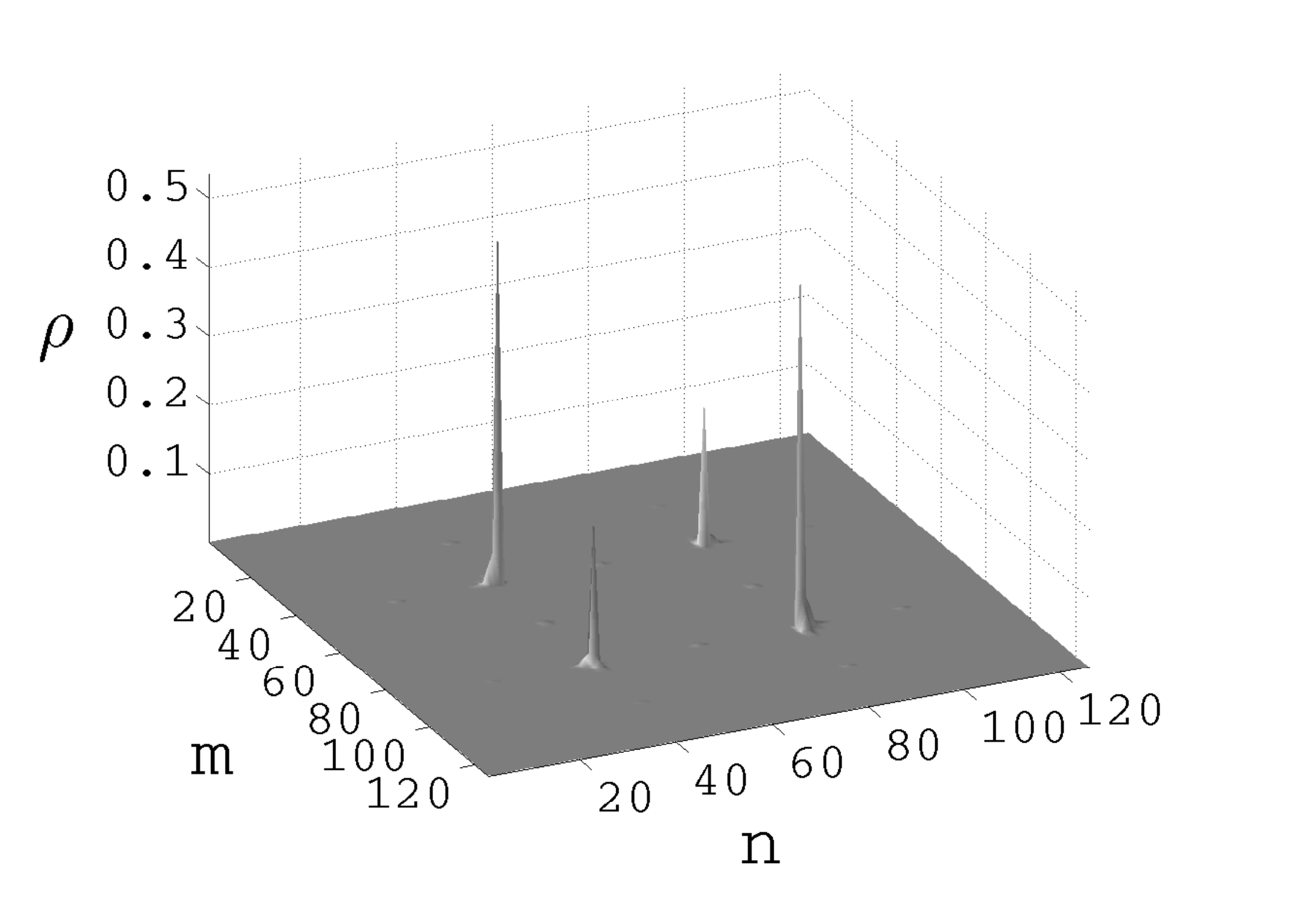}}\\
   \subfigure[]{\includegraphics[width=\textwidth/7*3]{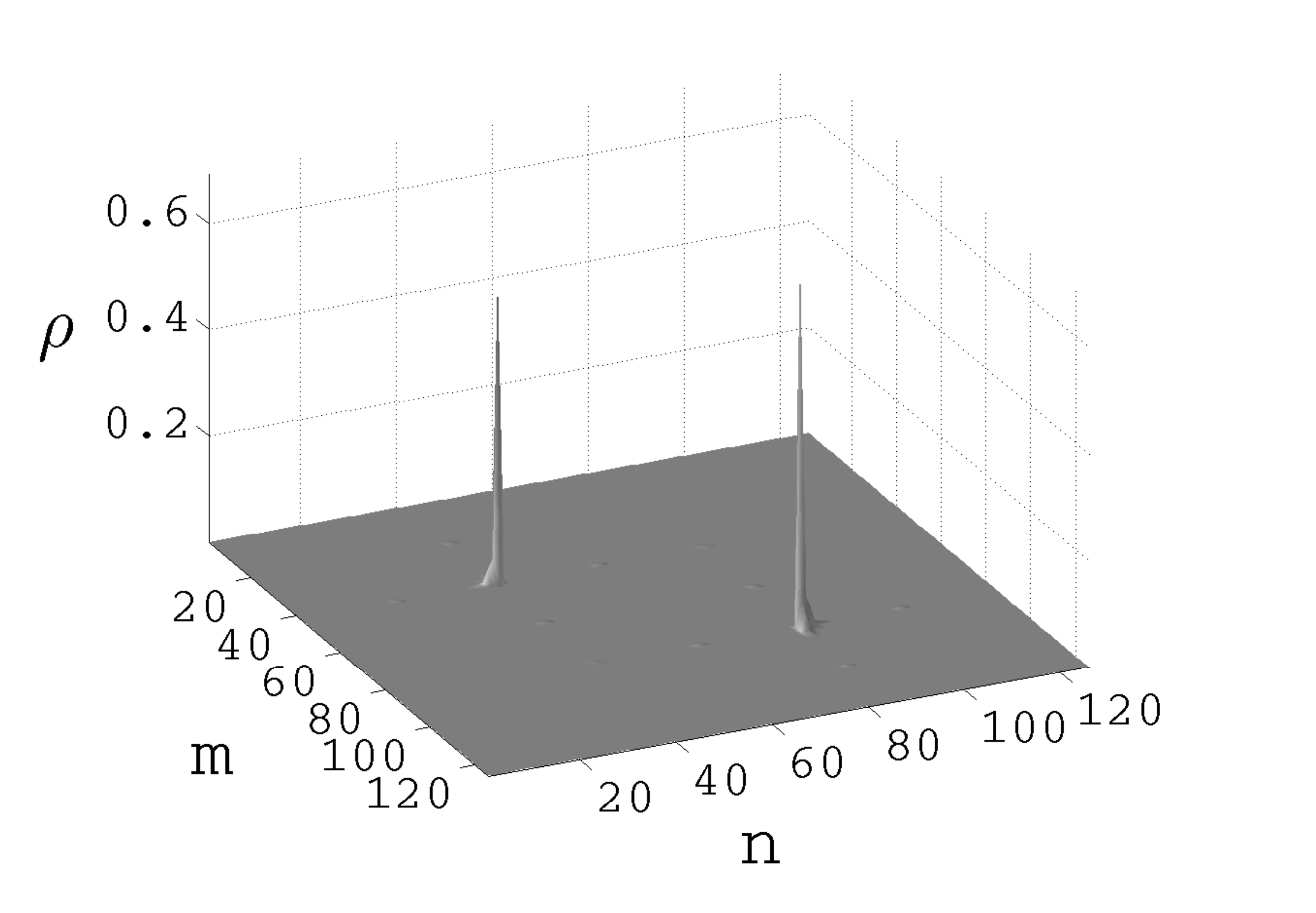}}\\
   \caption{\label{fig:rho} The magnitude of density-matrix at different times
        (61th excited state):
        (a) initial state, (b) $t=0.5$~sec, (c) $t=2$~sec.
	Fixed parameters: $a_{11}=2$~nm, $a_{12}=4$~nm, $a_{22}=3$nm, $N_1=5$, $N_2=20$,
        $\kappa_1=\kappa_2=0.005$.}
\end{figure}

Fig.~\ref{fig:rho} shows the magnitude of the density matrix of the 61th excited (cat) state at different times. At initial time $t=0$ there are two off-diagonal elements which are equal to the two diagonal ones.
While the diagonal elements do not change significantly in time, the off-diagonal ones fall off gradually, and the state becomes a classical mixture after 2 second.
A similar behavior (not shown) can be observed for the 63th cat state (and for all the other cat states).

\begin{figure}[hbt!]
   \subfigure[]{\includegraphics[width=\textwidth/8*3]{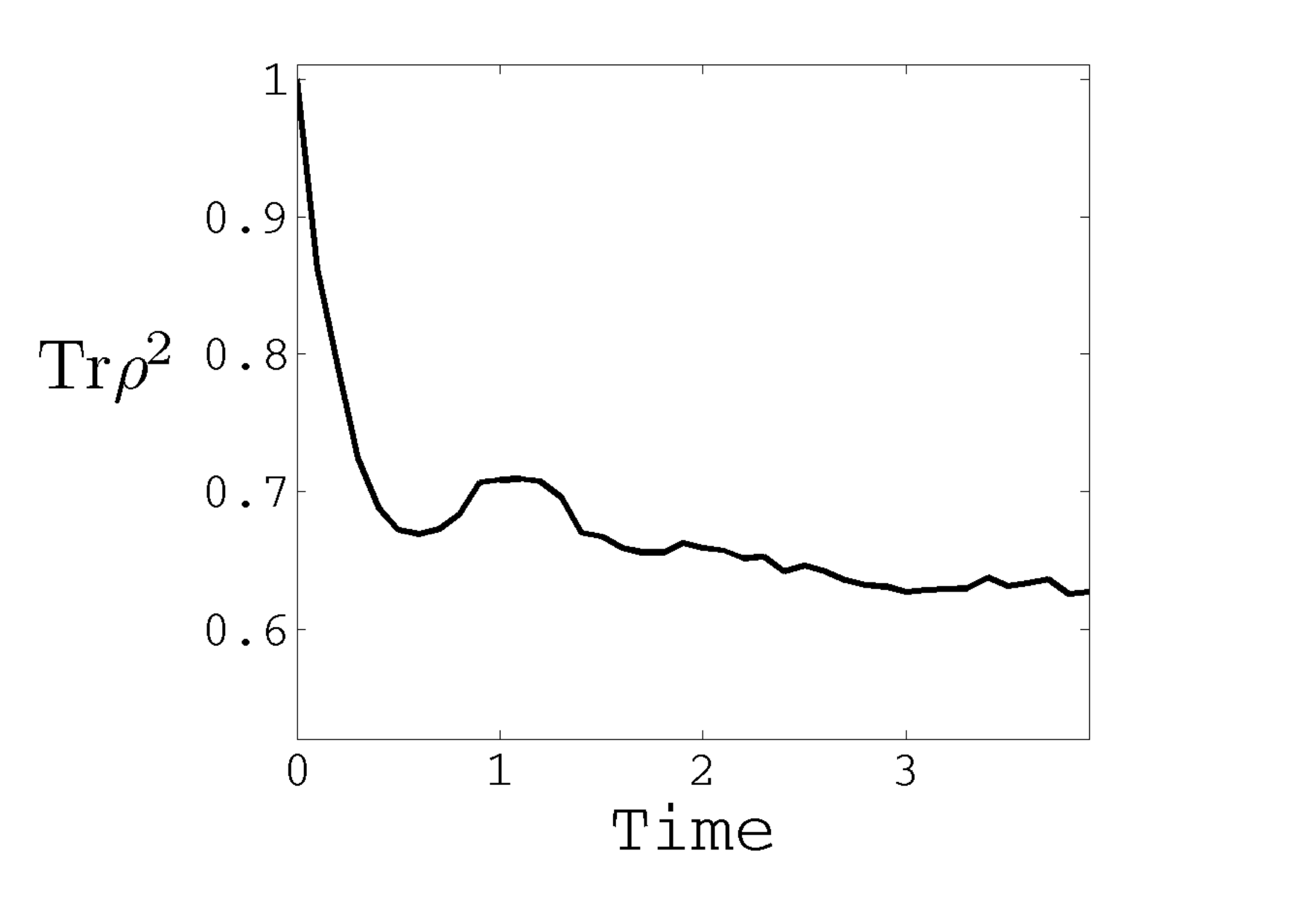}}\\
   \subfigure[]{\includegraphics[width=\textwidth/8*3]{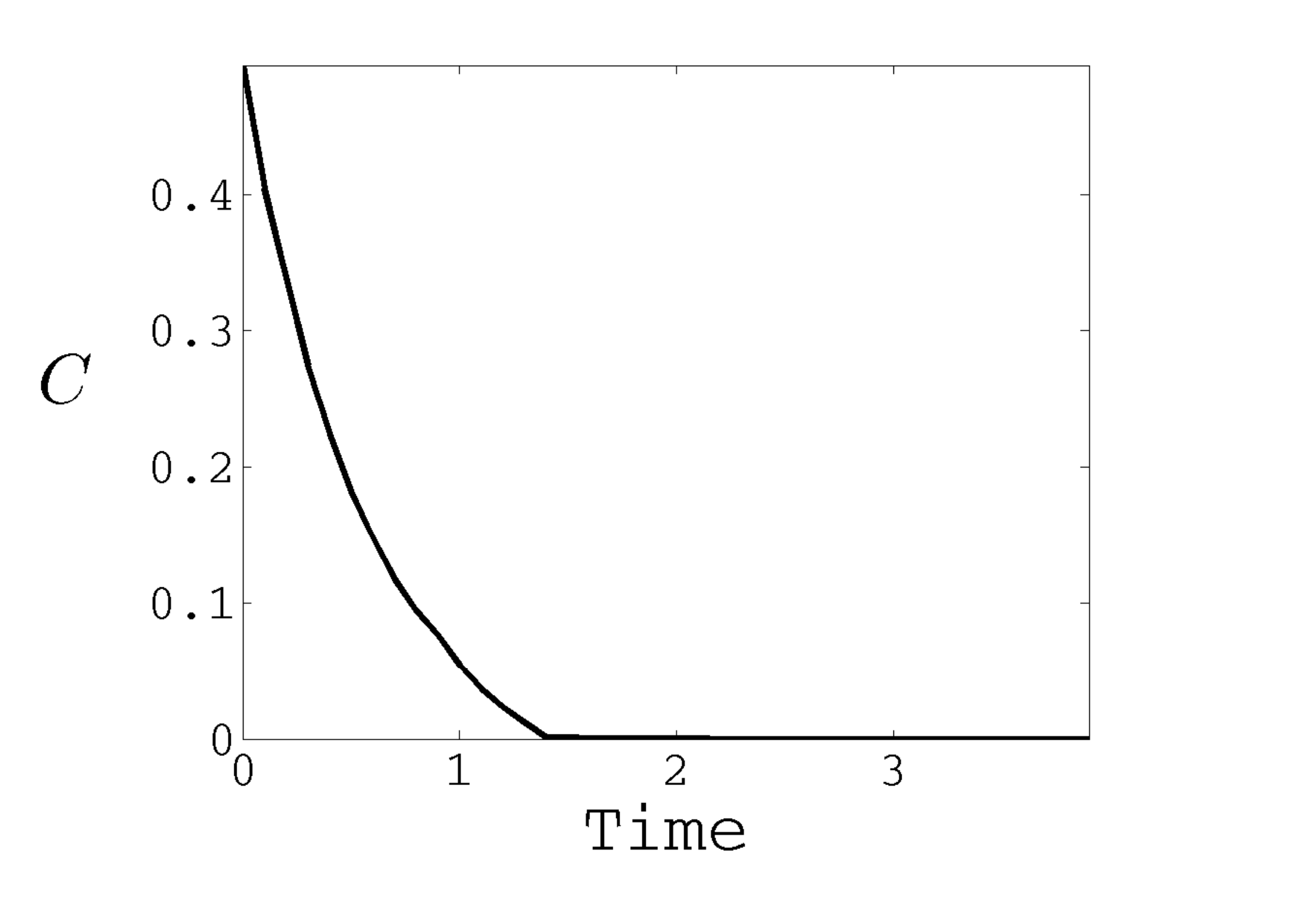}}\\
   \caption{\label{fig:state_61} The 61th excited state: $(\ket{1,4}_1\ket{20,0}_2+\ket{4,1}_1\ket{0,20}_2)/\sqrt{2}$.
	Fixed parameters: $a_{11}=2$~nm, $a_{12}=4$~nm, $a_{22}=3$nm, $N_1=5$, $N_2=20$,
        $\kappa_1=\kappa_2=0.005$. Time is measured in units of second.
        (a) $\mathrm{Tr}\rho^2$ as function of time.
        (b) The magnitude $C$ of the average of the off-diagonal elements as function of time.}
\end{figure}

To study finer details of the decoherence effect of environment on cat states we have calculated
both $\mathrm{Tr}\rho^2$  and the magnitude $C$ of the average of the two maximal off-diagonal elements
   as function of time for the selected
		61th and 63th excited states.
These two quantities are characteristic of time development of the decoherence process, measuring
formation of statistical mixture from the initital pure cat state.

\begin{figure}[hbt!]
   \subfigure[]{\includegraphics[width=\textwidth/8*3]{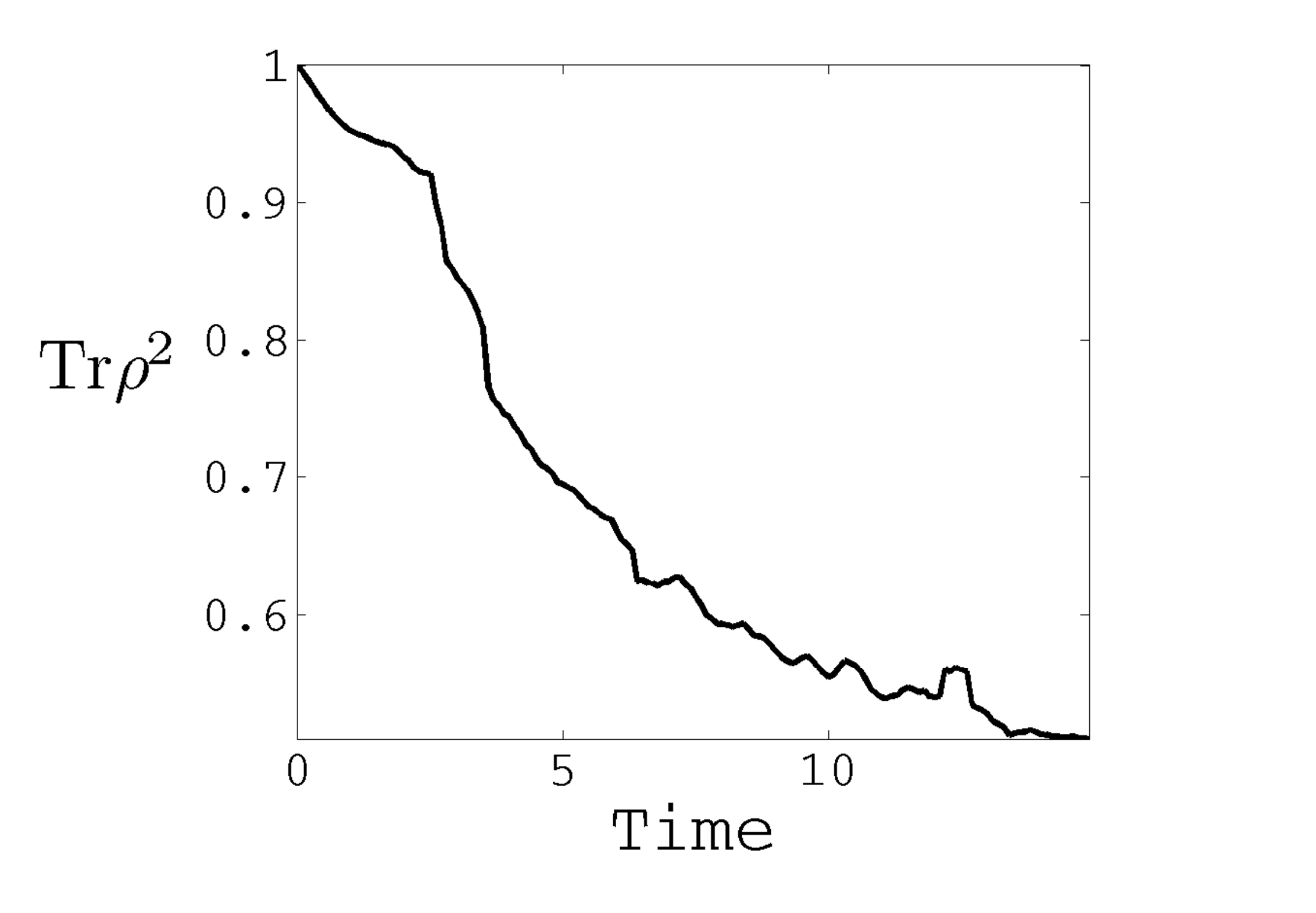}}\\
   \subfigure[]{\includegraphics[width=\textwidth/8*3]{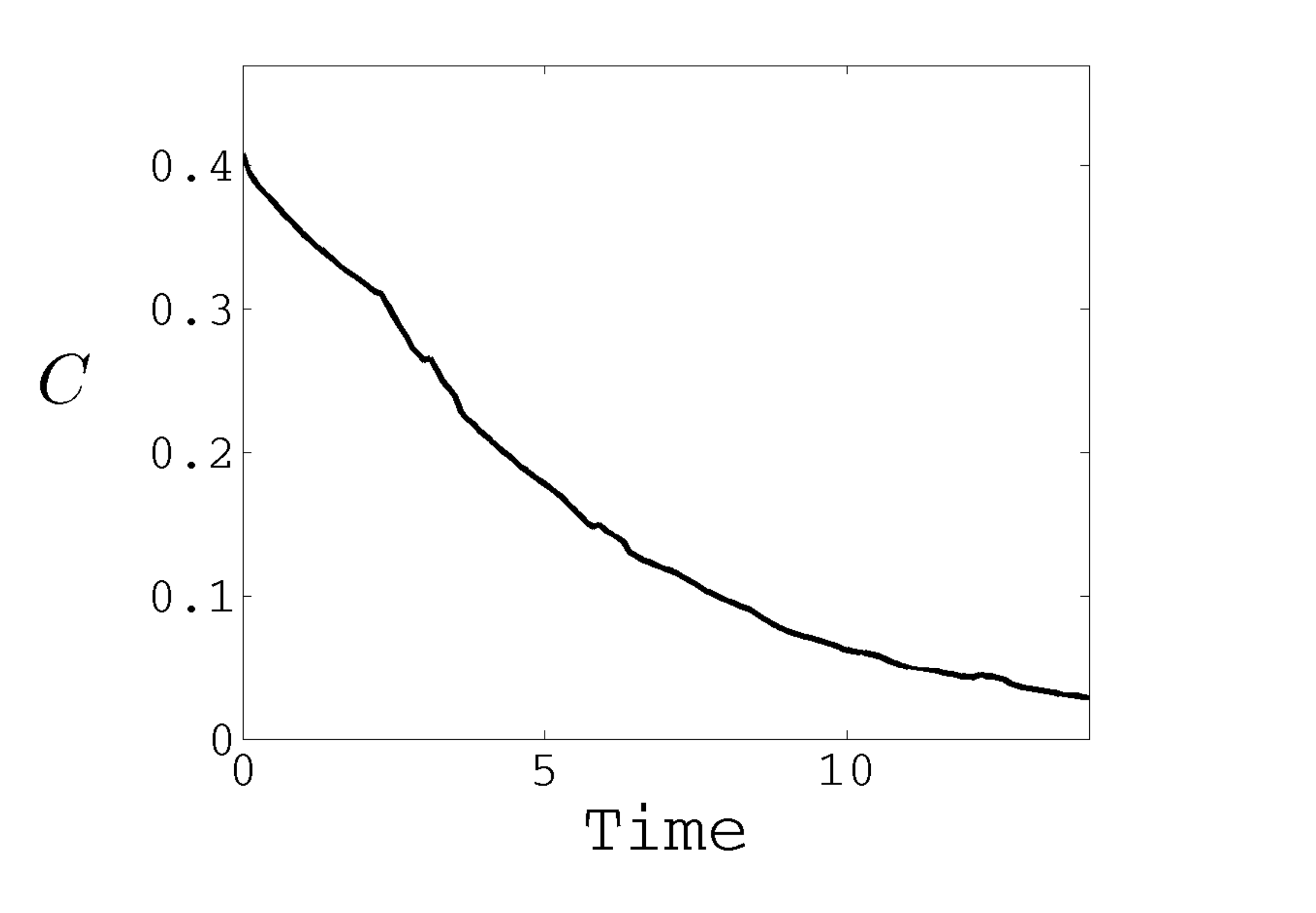}}\\
   \caption{\label{fig:state_63} The 63th excited state: $(\ket{2,3}_1\ket{7,13}_2+\ket{3,2}_1\ket{13,7}_2)/\sqrt{2}$.
	Fixed parameters: $a_{11}=2$~nm, $a_{12}=4$~nm, $a_{22}=3$nm, $N_1=5$, $N_2=20$,
        $\kappa_1=\kappa_2=0.005$. Time is measured in units of second.
        (a) $\mathrm{Tr}\rho^2$ as function of time.
        (b) The magnitude $C$ of the average of the off-diagonal elements as function of time.}
\end{figure}

The results are shown in Figs.~\ref{fig:state_61} and \ref{fig:state_63} and we can observe that the decoherence is more effective for the state where the difference $\Delta N_i=|N_i^L-N_i^R|$ of particle numbers of the two components $i=1,2$ in the left and right potential well is greater. For example, in case of the 61th excited cat state, $\Delta N_1=3$ and $\Delta N_2=20$ which is greater than the corresponding values $\Delta N_1=1$ and $\Delta N_2=6$ of the 63th state, and thus $\mathrm{Tr}\rho^2$ of the 61th state approaches more quickly the value 0.5, characteristic of the statistical mixture than that of the 63th state. A similar behavior can be observed when the time development of the average $C$ of the two maximal off-diagonal density matrix elements are studied in Figs.~\ref{fig:state_61} and \ref{fig:state_63}.

\section{Conclusion \label{sec:conclusion}}

We have studied macroscopic quantum superposition (cat) states  formed in two-component Bose-Einstein condensates described within the two-mode approximation. An analysis of the limitations of this approximation can be found in Ref.~\cite{Sakman2009}. However, the two-mode approximation has been ascertained to describe many aspects of double well  experiments~\cite{Albiez2005}.
The model has been applied to  BECs where the two components correspond to two internal hyperfine state of $^{87}$Rb atoms. Because of limited computational capacity, relatively modest particle numbers between 1-20 has been investigated, but a systematic check revealed that the main features of forming the cat states depend mostly on the actual interaction strengths (characterized by scattering lengths $a_{ij}$) between atoms, and not on the actual numbers $N_i, i=1,2$. Therefore we have performed exploratory calculations in wide range of values of scattering lengths $a_{ij}, i,j=1,2$, both in the attraction and repulsion domain, and presented the results in case of some selected examples exhibiting the most important features of formation, duration, and ceasing of MQSS or cat states.

We have found that, in  general, positive inter-species scattering lengths $a_{ii}, i=1,2$ enhance the probability $P=N_{cat}/N_{tot}$ of two-component cat state formation. Small or zero intra-species scattering lengths $a_{12}$, of course, lessen this probability, as the system of two-component BEC actually decouples into two independent one-component BECs where, however, formation of one-component cat states is possible. In case of positive inter-species scattering length $a_{12}$ also the individual number $N_{cat}$ of two-component cat states is growing as it can be seen in Fig.~\ref{fig:prob3b} but negative $a_{12}$ hinders the formation of cat state as Fig.~\ref{fig:prob4b} shows. Finally, in the case when all the scattering lengths take negative values, there is only a limited range of two-component BEC cat state formation, studied both in terms of probability $P$ or the individual cat state number $N_{cat}$, as Figs.~\ref{fig:prob5} and \ref{fig:prob5b} exhibit.

Effect of environment has been modeled by phase-damping which is the result of elastic interaction between condensed and non-condensed atoms and  preserves the particle numbers $N_1, N_2$. Using the corresponding Lindblad master equation \eqref{eq:master_eq} we have found that the most important parameter which gives rise to the decoherence effect of such type is the particle number difference $\Delta N_i=|N_i^L-N_i^R|, i=1,2$ between the left and right potential well. As Figs.~\ref{fig:state_61} and \ref{fig:state_63} suggest the decoherence  occurs more quickly for two-component cat states where the number difference $\Delta N_i$ is greater.

The influence of asymmetric wells, and other effects of the environment on the formation of cat states needs further research.

\section*{Acknowledgment}
The authors thank Dr. Daniel Schumayer for reading the manuscript and providing valuable comments.

\bibliographystyle{apsrev}
\bibliography{cat}

\end{document}